\documentclass[12pt,preprint]{aastex}
\usepackage{psfig}
\usepackage{epsfig}
\begin{document}

\title{{\em SPITZER} Observations of the SCUBA/VLA Sources in the
Lockman Hole: \\
Star Formation History of Infrared-Luminous Galaxies}

\author{E.\ Egami\altaffilmark{1}, 
H.\ Dole\altaffilmark{1,2}, 
J.-S.\ Huang\altaffilmark{3}, 
P.\ P\'{e}rez-Gonzalez\altaffilmark{1}, 
E.\ Le Floc'h\altaffilmark{1}, 
C.\ Papovich\altaffilmark{1}, 
P.\ Barmby\altaffilmark{3},
R.\ J.\ Ivison\altaffilmark{4,5}, 
S.\ Serjeant\altaffilmark{6}, 
A.\ Mortier\altaffilmark{6}, 
D.~T.\ Frayer\altaffilmark{7}, 
D.\ Rigopoulou\altaffilmark{8}, 
G.\ Lagache\altaffilmark{2},
G.~H.\ Rieke\altaffilmark{1},
S.~P.\ Willner\altaffilmark{3},
A.\ Alonso-Herrero\altaffilmark{9},
L.\ Bai\altaffilmark{1},
C.~W.\ Engelbracht\altaffilmark{1},
G.~G.\ Fazio\altaffilmark{3},
K.~D.\ Gordon\altaffilmark{1},
D.~C.\ Hines\altaffilmark{1},
K.~A.\ Misselt\altaffilmark{1},
S.\ Miyazaki\altaffilmark{10},
J.~E.\ Morrison\altaffilmark{1},
M.~J.\ Rieke\altaffilmark{1},
J.~R.\ Rigby\altaffilmark{1},
G.\ Wilson\altaffilmark{7}
}

\altaffiltext{1}{Steward Observatory, University of Arizona, 933
N. Cherry Avenue, Tucson, AZ85721}
\altaffiltext{2}{Institut d'Astrophysique Spatiale, Universit\'{e} Paris
Sud, Bat. 121, 91405, Orsay Dedex, France}
\altaffiltext{3}{Harvard-Smithsonian Center For Astrophysics, 60
Garden Street, Cambridge, MA02138}
\altaffiltext{4}{Astronomy Technology Centre, Royal Observatory,
Blackford Hill, Edinburgh EH9 3HJ, UK}
\altaffiltext{5}{Institute for Astronomy, University of Edinburgh,
Blackford Hill, Edinburgh EH9 3HJ, UK}
\altaffiltext{6}{Centre for Astrophysics \& Planetary Science, School
of Physical Sciences, University of Kent, Canterbury, Kent, CT2 7NR, UK}
\altaffiltext{7}{Spitzer Science Center, California Institute of
Technology, 314-6, Pasadena, CA~91125}
\altaffiltext{8}{Astrophysics, Denys Wilson Building, Keble Road,
Oxford, OX1, 3RH, UK}
\altaffiltext{9}{Consejo Superior de Investigaciones Cientificas,
Serrano, 117, 28006, Madrid, Spain}
\altaffiltext{10}{Subaru Telescope, National Astronomical Observaotry
of Japan, 650 North Aohoku Place, Hilo, HI96720}

%\dataset{ads/sa.spitzer\#6619904, 6620160}

\begin{abstract}

We present {\em Spitzer} IRAC (3.6, 4.5, 5.8, 8.0 $\mu$m) and MIPS
(24$\mu$m) observations of the SCUBA submillimeter sources and $\mu$Jy
VLA radio sources in a 5\arcmin$\times$5\arcmin\ area in the Lockman
Hole East region.  Out of the $\sim$40 SCUBA/VLA sources in the field,
{\em Spitzer} counterparts were detected for nearly all except for the
few low-weight SCUBA detections.  We show that the majority (80--90\%)
of the detected sources are cold (i.e., starburst-like)
infrared-luminous galaxies ($L_{IR} > 10^{11}$ L$_{\sun}$) at redshift
$0.5<z<3.5$, whose star-formation rate density (SFRD) is comparable to
that of the optically-selected star-forming galaxies.

\end{abstract}

\section{INTRODUCTION}

A major scientific goal of the {\em Spitzer Space Telescope} is to probe
the evolution of infrared-luminous galaxies ($L_{IR}>10^{11}$
L$_{\sun}$) at $z \ga 1$.  Neither IRAS nor ISO provided enough
sensitivity to detect a significant number of galaxies at $z>1$,
except for a small number of extreme objects.  The much improved
sensitivity and spatial resolution of {\em Spitzer} in the
mid/far-infrared have improved the situation dramatically.  With this
goal in mind, we have imaged a 5\arcmin$\times$5\arcmin\ area in the
Lockman Hole East region with InfraRed Array Camera (IRAC;
\citealt{Fazio04}) and Multiband Imaging Photometer for Spitzer (MIPS;
\citealt{Rieke04}).  The Lockman Hole was chosen because of the low
infrared cirrus background as well as the wealth of ancillary data.

Up to now, $z \ga 1$ infrared luminous galaxies have been probed
mainly through submillimeter \citep[e.g.,][]{Smail97} or radio
\citep[e.g.,][]{Barger00} observations.  The former directly detects
the emission by dust that is responsible for the infrared luminosity,
while the latter detects the synchrotron emission (presumably from
supernova remnants), which is known to correlate with the infrared
luminosity \citep[e.g.,][]{Helou85}.  These two types of selection
provide complementary views in that submillimeter selection is biased
toward ultra-luminous ($L_{IR} > 3-4 \times 10^{12}$L$_{\sun}$)
sources at high redshift (with a median redshift of 2.4
\citep{Chapman03}), while radio selection is biased toward less
luminous sources at intermediate redshift ($z=1-2$).

In this paper, we conduct the first analysis of SCUBA submillimeter
and VLA radio sources based on the new {\em Spitzer} data.  Companion papers
also discuss stacking analysis of faint SCUBA sources
\citep{Serjeant04}, MAMBO millimeter sources \citep{Ivison04a}, IRAC
sources \citep{Huang04}, X-ray sources \citep{Alonso04}, 24
$\mu$m-selected sources \citep{LeFloch04}, and extremely red objects
\citep{Wilson04} in the same field.

Throughout the paper, H$_{0}=70$ km s$^{-1}$ Mpc$^{-1}$,
$\Omega_{M}=0.3$, and $\Omega_{\Lambda}=0.7$ were assumed.
\section{Data and Sample}

\subsection{Observations and Data Reduction}

\begin{figure}
  \epsscale{0.9}
  \plotone{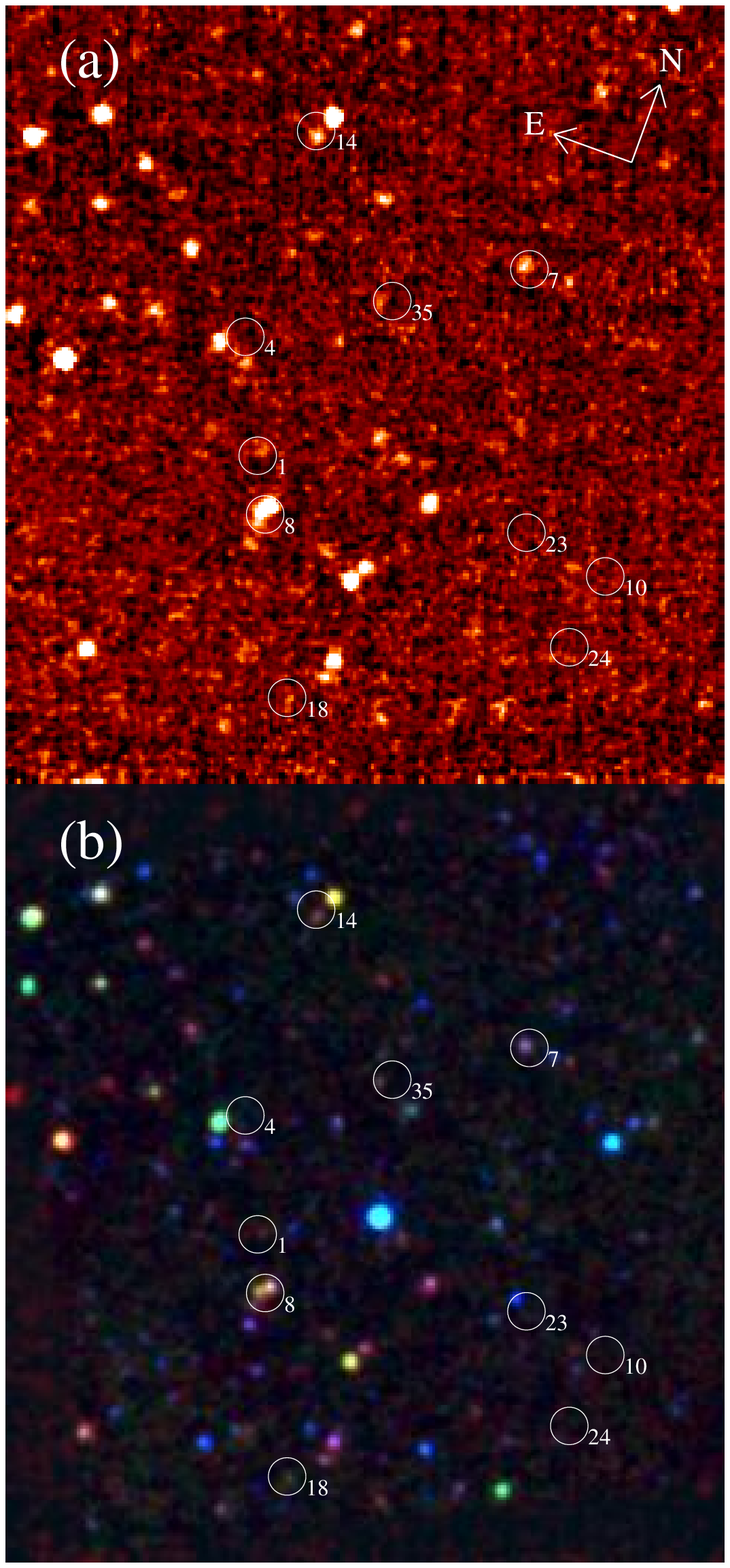}
  \caption[f1.eps]{(a) MIPS 24 $\mu$m image of the Lockman Hole East;
  (b) Three-color image of the same field produced from the IRAC 3.6
  $\mu$m, 8.0 $\mu$m, and MIPS 24 $\mu$m images.  The ten SCUBA
  sources (LE850.1/4/7/8/10/14/18/23/24/35) are indicated with white
  circles with a radius of 8\arcsec.  The field of view is
  $\sim$5\arcmin$\times$5\arcmin. \label{spitzer}}
\end{figure}

The details of the IRAC observations are presented by \citet{Huang04},
so only the MIPS observations are described here.  The MIPS 24~$\mu$m
images of the Lockman Hole East region were taken on UT 2003 November
30 (Program ID: 1077; AOR ID: 6619904).  The large-source photometry
mode was used to minimize the dither step lengths and maximize the
areal coverage overlap within the 5\farcm4 $\times$ 5\farcm4 MIPS
24~$\mu$m field of view.  The total integration time was 300 seconds
per pixel.  The 24~$\mu$m imaging data were reduced and combined with
the Data Analysis Tool (DAT) developed by the MIPS instrument team
\citep{Gordon04}.  The final reduced 24~$\mu$m image is shown in
Figure~\ref{spitzer}a while the three-color image produced from the
IRAC 3.6, 8, and MIPS 24~$\mu$m is shown in Figure~\ref{spitzer}b.
The 24~$\mu$m sources were extracted from the image as described in
\citet{Papovich04}.  The detection limit in the 24~$\mu$m image is
$\sim$ 120~$\mu$Jy (3 $\sigma$).  A preliminary IRAC/MIPS source
catalog of this field will be released at
\url{http://mips.as.arizona.edu/cs/ero}.

A variety of other imaging data were also assembled \citep{LeFloch04}.
This ancillary data set includes the F606W HST/WFPC2 images as well as
the $UBVRIJHK_{s}$ images obtained with a variety of ground-based
telescopes.  It also includes a deep (5 $\mu$Jy/beam) high-resolution
(1\farcs4 FWHM) VLA map~\citep{Ivison02} and a reprocessed SCUBA map
\citep{Mortier04}.

\subsection{The SCUBA/VLA Sources in the Field}

The field imaged by both IRAC and MIPS contains 10 SCUBA sources in
the 850 $\mu$m catalog of \citet{Scott02},
LE850.1/4/7/8/10/14/18/23/24/35\footnote{The source names are based on
\citet{Ivison02}.  The full names according to SIMBAD are of the form
[SFD2002]LHE N.} (Figure~\ref{spitzer}).  The first seven sources are
considered secure while the last three are regarded as marginal, the
distinction being whether the signal-to-noise ratio is above or below
3.5.  Among the seven secure sources, \citet{Ivison02} detected radio
counterparts for five (LE850.1/7/8/14/18) and possibly for one more
(LE850.4) with a lower significance. Our re-examination of the VLA map
shows that LE850.35 also has a VLA source within a radius of 8\arcsec,
the region of 95 \% positional confidence according to
\citet{Ivison02}.  We tentatively identify this VLA source as the
counterpart for LE850.35 although we could not confirm the reality of
this submillimeter source unambiguously even with the reprocessed
SCUBA map.  LE850.35 makes the total of radio-detected SCUBA sources
seven out of ten. The field also contains a total of 38 VLA sources
above 20 $\mu$Jy (4$\sigma$ peak flux density) using the map of
\citet{Ivison02}.

\section{RESULTS and DISCUSSION}

\subsection{Identification of SCUBA Sources}

\begin{figure}
  \epsscale{0.8}
%  \centerline{\bf See the JPEG file f2.jpg.}
  \plotone{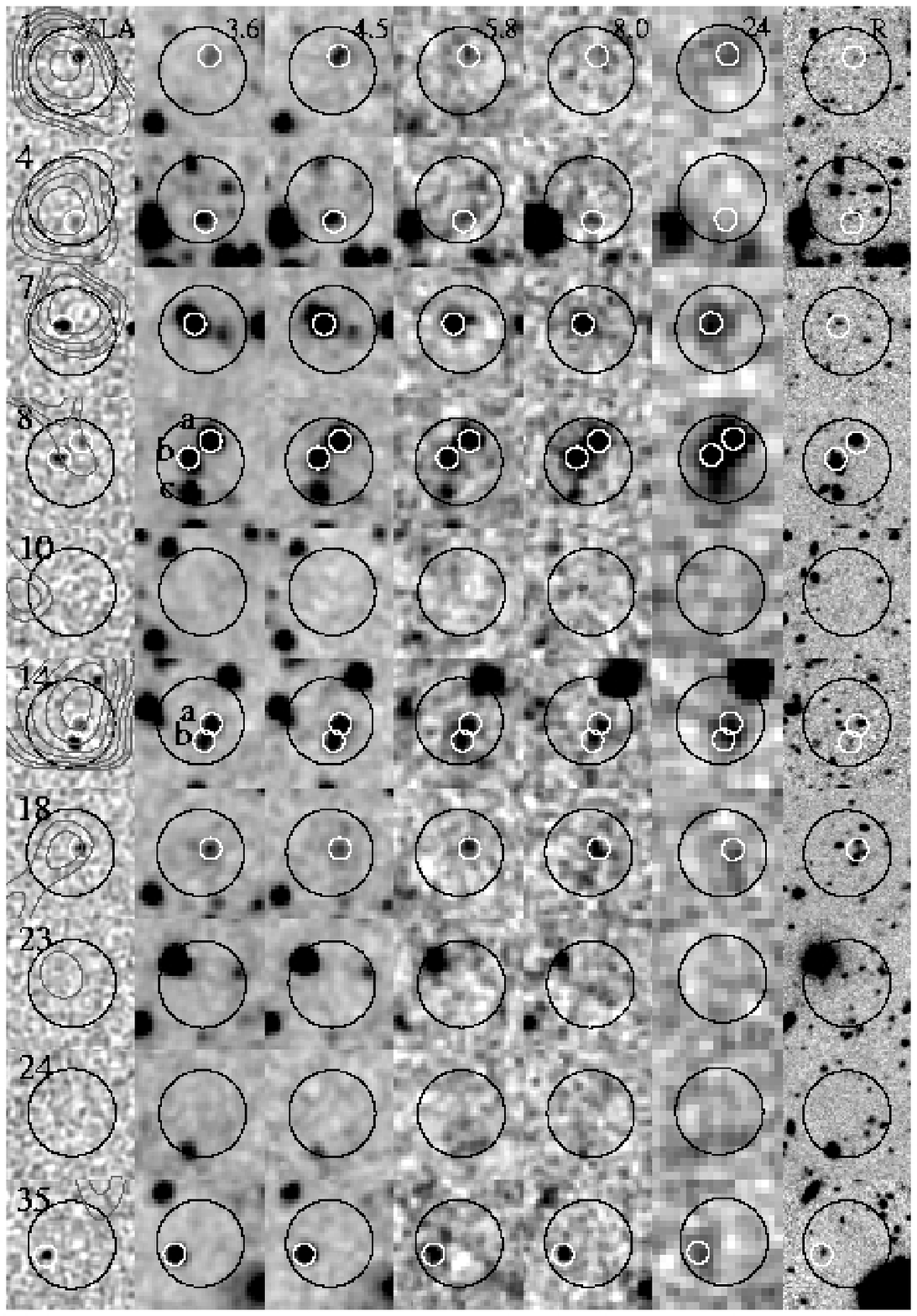}
  \caption[f2.eps]{Postage stamp images of the ten SCUBA sources.
  Shown here are (from left to right) VLA, IRAC 3.6, 4.5, 5.8, 8.0
  $\mu$m, MIPS 24 $\mu$m, and R-band images.  The VLA images are from
  the data of \citet{Ivison02} while the R-band images are from the
  data taken from the Subaru telescope archive.  The 850 $\mu$m
  contour maps \citep{Mortier04} are overlaid on the VLA images.  The
  contours start at 2.5$\sigma$ above the sky and increases by a factor
  of 1.2.  The size of each image is 20\arcsec$\times$20\arcsec.  The
  solid circle indicates the region of 95 \% positional confidence
  ($r=8$\arcsec) by \citet{Ivison02}.  The small white circles
  indicate the positions of the radio sources seen in the VLA map. In
  all the images, north is up and east is left.
  \label{poststamp}}
\end{figure}

Figure~\ref{poststamp} shows the {\em Spitzer} as well as VLA and
archived Subaru/Suprime-Cam \citep{Miyazaki02} $R$-band images of all
ten SCUBA sources in the field.  The seven sources with radio
detections have all been detected by IRAC while six (excepting only
LE850.4) have been detected by MIPS.  The IRAC/MIPS flux densities of
these sources are listed in Table~\ref{flux} together with the
previously published 850 $\mu$m and 20cm flux densities.  The
following three sources have more than one IRAC source within a radius
of 8\arcsec:

LE850.7 --- The IRAC 3.6 and 4.5 $\mu$m images show extended emission
north-east from the radio position, which is seen to be a nearby
source in the $R$-band image.  However, the longer-wavelength images
do not show emission in the same direction, indicating that this
neighboring source does not contribute significantly above $\sim$ 5
$\mu$m.

LE850.8 --- Three IRAC sources (8a/8b/8c) are seen within a
radius of 8\arcsec.  The 24 $\mu$m signal is clearly extended in a way
consistent with all three sources contributing.  The
strongest radio source corresponds to component 8b.  There is also 
faint (3 $\sigma$) radio emission at the position of component 8a,
which is the brightest source at 24 $\mu$m.  Component 8a was thought
to be the counterpart for the ROSAT X-ray source with a spectroscopic
redshift of 0.974 \citep{Lehmann01}, but the later XMM-Newton
observation showed that the X-ray source is actually coincident with
component 8b \citep{Ivison02}.  This leaves a possibility that some of
the radio flux from component 8b may be of AGN origin.  The $R$-band
image shows several sources at the position of component 8b.

LE850.14 --- The two radio counterparts (14a/14b) have both been
detected by IRAC.  The 24 $\mu$m emission is centered between the two
IRAC sources.  Considering the equally strong radio fluxes,
both components are expected to contribute significantly in the
submillimeter.  

In summary, we have 7 secure SCUBA sources with a total of 9 radio
components (LE850.1/4/7/8a/8b/14a/14b/18/35), all of which have {\em
Spitzer} counterparts.  {\em Spitzer} counterparts have been found
only for the sources with radio detections.  The three sources which
are not detected by VLA and {\em Spitzer} (LE850.10/23/24) are all
weak in the reprocessed 850 $\mu$m map (Figure~\ref{poststamp}).

\subsection{Spectral Energy Distribution and Redshifts}

For the following statistical analysis, we restrict our discussion to
the VLA 20 $\mu$Jy sample, which contains 7 of the 9 SCUBA radio
components (LE850.1/7/8b/14a/14b/18/35).  There is one VLA source
which is relatively bright in the radio (230 $\mu$Jy) but undetected
in most other bands including the IRAC 5.8 and 8.0 $\mu$m (this is the
only source which has not been detected in all four IRAC bands).  We
exclude this source; The VLA sample consists of 37 objects, of which
all are detected in the four IRAC bands, and 29 are detected at 24
$\mu$m.

The SEDs of the VLA sources can be classified into two types: those
showing a clear near-infrared stellar continuum hump around 1.6 $\mu$m
(32 sources, 86\%) and those showing a featureless power-law continuum
(5 sources, 14\%).  Photometric redshifts can be derived for the
former using the near-infrared hump
\citep[e.g.,][]{Sawicki02,LeFloch04}; For the latter, the redshifts
are mainly constrained by the submillimeter and radio points, but the
estimates are significantly more uncertain.

\begin{figure}
  \epsscale{0.7}
  \plotone{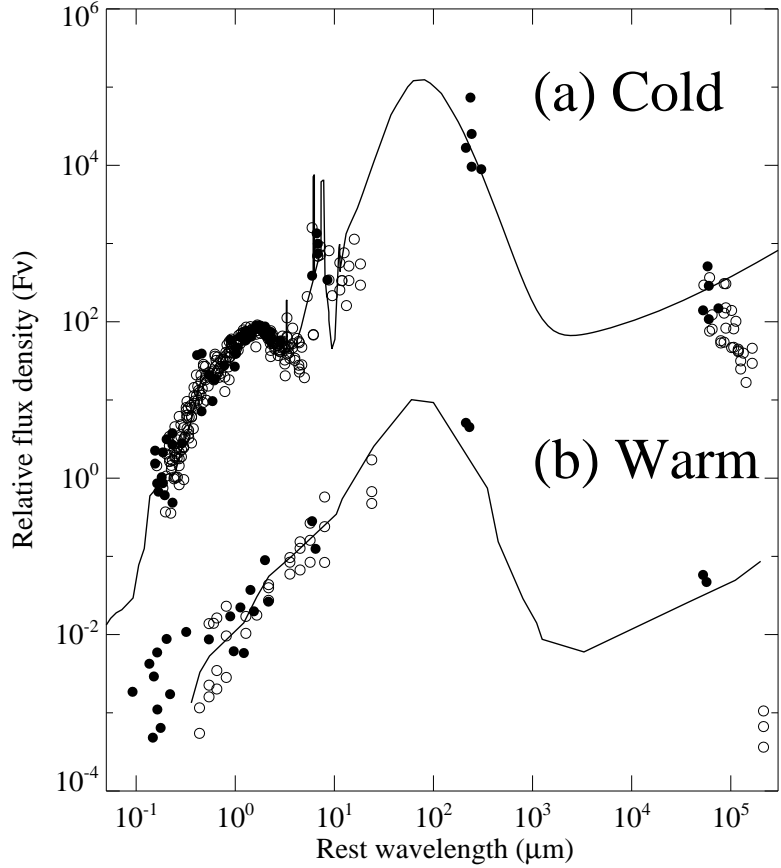}
  \caption[f3.ps]{The SEDs of 37 VLA sources detected above 20
  $\mu$Jy, which contain 7 SCUBA sources: (a) The composite SED of 32
  sources with a near-infrared continuum hump.  The SED is well fitted
  by that of a cold ULIRG Arp~220 (solid line; the model by
  \citet{Silva98}); (b) The composite SED of 5 power-law continuum
  sources.  The SED resembles that of a warm ULIRG Mrk~231 (solid
  line; the observed SED).  The solid circles denote SCUBA sources
  while the open circles denote VLA sources without SCUBA detections.
  The SEDs are shifted to the restframe using
  photometric/spectroscopic redshifts, and normalized around 1.6
  $\mu$m (i.e., the peak of the stellar continuum) with the cold
  sources and at 1--10 $\mu$m with the warm sources.  For the three
  warm VLA sources without submillimeter detections, for which
  redshift cannot be estimated, zero redshift was used.
  \label{sed}}
\end{figure}

Figure~\ref{sed} shows the composite SEDs of these two types.  The SED
of the sources with a near-infrared continuum hump is similar to that
of a ``cold'' ultra-luminous infrared galaxy (ULIRG) like Arp~220
while the SED of the power-law continuum sources resembles that of a
``warm'' ULIRG like Mrk~231.  This cold/warm classification often used
with ULIRGs roughly separates starburst-dominated galaxies (cold) and
AGN-dominated ones (warm) by the appearance of the infrared SEDs.
Among the 32 cold-type sources, there are 8 sources (22\% of the 37
VLA sample) which were not detected at 24 $\mu$m, but their restframe
near-infrared SEDs are similar to those of the 24 $\mu$m detected
sources.  Considering that these sources have expected 24 $\mu$m
fluxes (based on the Arp~220 SED) close to or below our detection
limit, we include them in the cold category.

As Figure~\ref{sed}a shows, the cold-type sources show remarkably
similar SEDs from the restframe visible to near-infrared that are
well-fitted by the SED of Arp~220.  Such uniformity is in contrast to
the diversity of the SEDs seen with the X-ray and 24 $\mu$m selected
sources \citep{Alonso04,LeFloch04}.  Among the SCUBA sources,
LE850.1/7/14a/14b/35 have this type of SEDs.  (LE850.4 is also of this
type, but it is not formally in the radio-selected sample.)  With the
SCUBA sources, the good fit to the Arp~220 SED extends to
submillimeter and radio while most of the VLA-only sources have
significantly lower radio fluxes, which reflects the fact that they are
not as infrared-luminous as Arp~220.  Figure~\ref{sed}a also shows that
higher redshift sources (i.e., those with radio points at shorter
wavelengths) have radio SED points closer to that of Arp~220.

The small SED dispersion with the cold-type sources demonstrates not
only the similarity of the SEDs but also the accuracy of the
photometric redshifts based on the near-infrared continuum hump.
Table~\ref{flux} shows that the redshifts derived this way for the
SCUBA sources are consistent with those based on the submm/radio flux
ratio \citep{Aretxaga03}.  LE850.14a and 14b show very similar SEDs,
which indicates that their redshifts are both $z\sim2.5$.  In the case
of LE850.14a, this was confirmed by the spectroscopic redshift of 2.38
\citep{Ivison04b}.  This suggests a possibility that these two sources
are physically associated.

If we regard the warm sources as AGN galaxies, the AGN fraction of the
20 $\mu$Jy radio sample is 14\% (5 out of 37).  Among the 9 radio
components associated with the SCUBA sources, the AGN fraction is
$\sim$ 30\% (3 out of 9) as shown in Table~\ref{flux}.  A comparison
with the 150~ksec XMM data \citep{Alonso04} shows that only 2 out of
the 5 power-law continuum sources are detected, and that there are
three cold sources detected in the X-ray.

Figure~\ref{vla}a shows the redshift distribution of the 32 cold
sources.  Assuming that these are starburst-dominated, the figure can
be regarded as the redshift distribution of infrared-luminous
star-forming galaxies selected in the radio.  The radio-infrared
correlation suggests that the infrared luminosities range from
$10^{11}$ to $10^{13}$ L$_{\sun}$, except for a few sources at
$z<0.5$, which are below $10^{11}$ L$_{\sun}$.  The SCUBA sample
constitutes only a small subset of the radio sample at high redshift.
This is because (1) due to the relatively high detection limit ($L_{IR}
> 3-4 \times 10^{12}$ L$_{\sun}$), the submillimeter observation
misses lower luminosity sources which radio observation can detect at
$z\la3$ \citep[e.g.,][]{Barger00}, and (2) because of the strong
luminosity evolution of the infrared-luminous galaxies, large
luminosity systems detectable above the submillimeter detection limit
are more abundant at higher redshift \citep[e.g.,][]{Smail97}.

\subsection{Star Formation History}

\begin{figure}
  \epsscale{0.6} \plotone{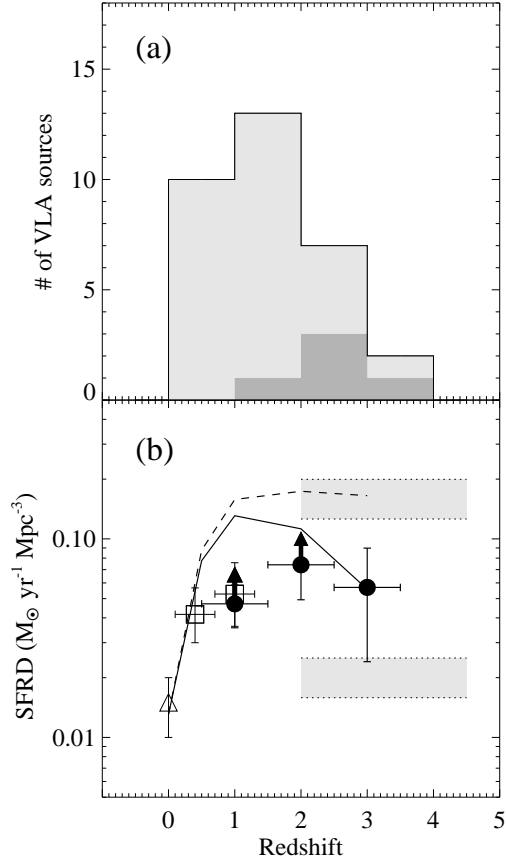}
  \caption[f4.eps]{(a) The redshift distribution of 32 cold-type VLA
  sources.  The dark shaded region indicates the 5 SCUBA sources.  (b)
  Star formation rate density (SFRD) of the cold-type galaxies plotted
  as a function of redshift with three redshift bins (0.5--1.5,
  1.5--2.5, 2.5--3.5).  The bins contain 17, 9, and 3 sources,
  respectively, and the resultant SFRDs are plotted as solid circles
  with the error bars based on the Poisson counting uncertainty (i.e.,
  SFRD divided by $\sqrt{N}$).  The open squares are from
  \citet{Barger00}, and the open triangle is from \citet{Yun01}.  No
  completeness correction was made for the $z>0$ points.  The lower
  shaded band indicates the range of observed SFRD with optically
  selected galaxies while the upper shaded band indicates the range
  with dust extinction correction
  \citep[e.g.,][]{Steidel99,Giavalisco04,Bouwens04}.  The dashed line
  indicates the total SFRD predicted by the model of \citet{Lagache04}
  while the solid line shows the model SFRD measurable above the radio
  flux limit of 20 $\mu$Jy.  The arrows indicate the increase of SFRD
  when we include the 24 $\mu$m selected infrared-luminous sources
  from \citet{LeFloch04} which have been missed by the radio observation.
  \label{vla}}
\end{figure}

Using the derived redshift distribution and converting the 1.4 GHz
radio luminosity into a star formation rate with the conversion factor
of \citet{Yun01}, we have calculated the star formation rate density
(SFRD) as a function of redshift (Figure~\ref{vla}b).  A K-correction
was applied based on a $\nu^{-0.8}$ spectrum.  Our SFRD at $z=1$ is
very close to the value derived by \citet{Barger00} with a sample of
$\mu$Jy VLA sources with spectroscopic redshifts.

The observed SFRD of the VLA sample shows a small peak at $z=2$, but
otherwise stays flat at $z=1-3$ above the observed SFRD of optically
selected star-forming galaxies.  When corrected for dust extinction,
the latter exceeds the former by a factor of a few.  However, the SFRD
of the infrared-luminous sample includes only the most extreme
high-luminosity objects at high redshift.  A substantial completeness
correction is still required, which has already been applied to the
optical sample.

To estimate the size of the completeness correction, the observed SFRD
was compared with a model prediction by \citet{Lagache04}.  This model
incorporates a variety of the latest observational constraints
including the MIPS 24 $\mu$m number counts \citep{Papovich04}.
However, the behavior of the model SFRD in Figure~\ref{vla}b is not
unique to this model and is similar to what was derived previously
\citep[e.g.,][]{Frances01}.  The observed points (solid circles) and
the model-predicted SFRD adjusted for the 20 $\mu$Jy radio detection
limit (solid line) are seen to agree at $z=2-3$, especially when we
include the 24 $\mu$m detected infrared-luminous galaxies which should
have been in the radio sample (based on the 24 $\mu$m $\rightarrow$ IR
$\rightarrow$ radio luminosity conversion) but are not (e.g., sources
larger than a few arcseconds would have been resolved away in the
radio map).  This suggests that if the completeness correction is
applied, the total SFRD of this infrared-luminous galaxy population,
which is considered to be distinct from the optically selected one
\citep[e.g.,][]{Frances01}, is flat at $z=2-3$ at a level comparable
to the SFRD of the optically selected galaxies.  At $z\sim1$, however,
the model overpredicts the observed SFRD by almost a factor of two.
We will examine in the future papers whether this discrepancy is
particular to this field (e.g., cosmic variance) or persistent even
with a larger sample.

\section{Summary}

Except for the few low-weight SCUBA detections, nearly all the SCUBA
and VLA $\mu$Jy sources have been detected by {\em Spitzer}.  The
majority (86\%) of the sources show remarkably similar SEDs well
fitted by the SED of a cold ULIRG like Arp~220.  This characteristic
was exploited to derive the redshift distribution, which then enabled
us to derive the star formation history of this galaxy population
using the radio-infrared correlation.  These submillimeter/radio
selected galaxies are mostly star-forming (i.e., cold)
infrared-luminous galaxies ($L_{IR}>10^{11}$ L$_{\sun}$) at $0.5<z<3.5$,
whose SFRD is probably comparable to that of the optically-selected
star-forming galaxies.

\acknowledgments

We would like to thank S.~C. Chapman for communicating us the revised
redshift of LE~850.18 before publication, and G. Neugebauer for
commenting on the manuscript.  This work is based in part on
observations made with the Spitzer Space Telescope, which is operated
by the Jet Propulsion Laboratory, California Institute of Technology
under NASA contract 1407. Support for this work was provided by NASA
through Contract Number 960785 issued by JPL/Caltech.

\clearpage

\clearpage

\begin{deluxetable}{lrrrrrrrcccc}
\tabletypesize{\footnotesize}
\rotate
\tablewidth{0pt}
\tablecaption{SPITZER flux measurements of the SCUBA sources \label{flux}}
\tablehead{
\colhead{Source} & \colhead{3.6 $\mu$m} & \colhead{4.5 $\mu$m} & \colhead{5.8 $\mu$m} & \colhead{8.0 $\mu$m} &
\colhead{24 $\mu$m} & \colhead{850 $\mu$m} & \colhead{20 cm} &
\colhead{$z_{phot}$\tablenotemark{a}} & \colhead{$z_{spec}$\tablenotemark{b}} & \colhead{$z_{phot}$} & \colhead{Type} \\
\colhead{(LE850)}       & \colhead{($\mu$Jy)}  & \colhead{($\mu$Jy)}  & \colhead{($\mu$Jy)}  & \colhead{($\mu$Jy)}  & 
\colhead{($\mu$Jy)} & \colhead{(mJy)}      & \colhead{($\mu$Jy)} &
\colhead{(submm/radio)} & \colhead{} & \colhead{(this work)} &\colhead{}  }
\startdata
1   &  3.8$\pm$0.6 &  8.2$\pm$1.1 & 11.8$\pm$2.5 &  10.2$\pm$1.7  & 193$\pm$33  & 10.5$\pm$1.6  &  73$\pm$10 & 2.6$\pm_{0.5}^{0.4}$ & \nodata & 2.6       & cold \\
4   &  7.7$\pm$1.0 & 10.2$\pm$1.4 & 11.5$\pm$2.5 &   9.6$\pm$1.6  &   \nodata   &  8.3$\pm$1.8  &  19$\pm$8  & 4.7$\pm_{0.5}^{1.3}$ & \nodata & 2.6       & cold \\
7   & 65.9$\pm$6.9 & 77.0$\pm$8.1 & 60.1$\pm$7.8 &  52.4$\pm$5.9  & 312$\pm$39  &  8.1$\pm$1.9  & 135$\pm$13 & 3.6$\pm_{1.9}^{0.4}$ & \nodata & 1.8       & cold \\
8b  & 17.1$\pm$2.0 & 22.3$\pm$2.6 & 37.2$\pm$5.3 &  89.5$\pm$9.6  & 282$\pm$59  &  5.1$\pm$1.3  &  58$\pm$12 & 3.7$\pm_{0.7}^{1.5}$ & \nodata & $\sim$3   & warm \\
8a  & 72.0$\pm$7.5 & 72.6$\pm$7.6 & 79.0$\pm$9.7 & 112.0$\pm$11.9 & 534$\pm$117 & \nodata       &  22$\pm$11 & \nodata  & 0.974   & \nodata   & warm \\
8c  & 26.7$\pm$3.0 & 20.7$\pm$2.4 & 17.2$\pm$3.1 &  17.8$\pm$2.4  & 161$\pm$47  & \nodata       &  \nodata   & \nodata  & \nodata & 0.9       & cold \\
14b &  9.8$\pm$1.3 & 14.7$\pm$1.8 & 22.2$\pm$3.7 &  18.5$\pm$2.5  & 166\tablenotemark{c} &  6.3\tablenotemark{c} &  72$\pm$12 & 2.4$\pm_{0.4}^{1.9}$ & \nodata & 2.5 & cold \\
14a & 15.2$\pm$1.9 & 20.2$\pm$2.4 & 27.6$\pm$4.3 &  18.6$\pm$2.5  &  83\tablenotemark{c} &  3.2\tablenotemark{c} &  36$\pm$12 & \nodata  & 2.38    & 2.5 & cold \\ 
18  &  6.1$\pm$0.9 &  5.8$\pm$0.9 & 19.9$\pm$3.4 &  26.4$\pm$3.3  & 125$\pm$33  &  4.5$\pm$1.3  &  47$\pm$10 & 3.0$\pm_{1.0}^{1.1}$ & 2.69    & $\sim$3   & warm \\
35  & 23.5$\pm$2.7 & 25.9$\pm$2.9 & 26.3$\pm$4.1 &  34.8$\pm$4.2  & 155$\pm$31  &  6.7$\pm$2.3  &  56$\pm$10 & \nodata  & \nodata & 3.0       & cold 
\enddata

\tablecomments{The photometric uncertainties include the 10\%
  calibration errors.  The 850 $\mu$m fluxes and 20 cm fluxes are from
  \citet{Scott02} and \citet{Ivison02}.}
\tablenotetext{a}{Photometric redshifts (le2 type) from \citet{Aretxaga03}}
\tablenotetext{b}{Spectroscopic redshifts for LE850.8a \citep{Lehmann01}, LE850.14a \citep{Ivison04b}, and LE850.18 \citep{Chapman04}.}
\tablenotetext{c}{The measured total flux densities of LE850.14 at 24 $\mu$m
  (249$\pm$39 $\mu$Jy) and 850 $\mu$m (9.5$\pm$2.8 mJy) were distributed based on the 20cm flux density ratio.}

\end{deluxetable}

\end{document}